\documentstyle[prl,aps,epsf,epsfig,multicol]{revtex}

\newcommand{\be}{\begin{equation}}
\newcommand{\ee}{\end{equation}}
\newcommand{\bes}{\begin{eqnarray}}
\newcommand{\ees}{\end{eqnarray}}
\newcommand{\bma}{\left( \begin {array}}
\newcommand{\ema}{\end {array} \right)}

\begin{document}

\title{2-d Self-Avoiding Walks on a Cylinder}

\author{Helge Frauenkron, Maria Serena Causo, and Peter Grassberger}

\address{HLRZ c/o Forschungszentrum J\"ulich, D-52425 J\"ulich, Germany}

\date{\today}

\maketitle
\begin{abstract}  
  We present simulations of self-avoiding random walks on 2-d lattices with
  the topology of an infinitely long cylinder, in the limit where the
  cylinder circumference $L$ is much smaller than the Flory radius. We
  study in particular the $L$-dependence of the size $h$ parallel to the
  cylinder axis, the connectivity constant $\mu$, the variance of the
  winding number around the cylinder, and the density of parallel contacts.
  While $\mu(L)$ and $\langle W^2(L,h)\rangle$ scale as as expected (in
  particular, $\langle W^2(L,h)\rangle \sim h/L$), the number of parallel
  contacts decays as $h/L^{1.92}$, in striking contrast to recent
  predictions. These findings strongly speak against recent speculations
  that the critical exponent $\gamma$ of SAW's might be nonuniversal.
  Finally, we find that the amplitude for $\langle W^2\rangle$ does not
  agree with naive expectations from conformal invariance.

  \vspace{4pt}
\end{abstract}

\begin{multicols}{2}
Although self-avoiding random walks (SAW's) are among the best studied 
critical phenomena, there have been recently speculations which, if true, 
would have very far reaching and surprising consequences \cite{cardy}. One such 
consequence would be that the critical exponent $\gamma$ of interacting 
SAW's on the square lattice would be 
temperature dependent if the interaction between neighboring 
bonds (i.e. steps of the walk on opposite edges of a plaquette) depends 
on the relative orientation of the steps. Another consequence would be 
that $\gamma$ for a-thermal SAW's on the Manhattan lattice is different 
from $\gamma$ on other 2-d lattices \cite{bennett}. Numerical evidence 
for the former prediction is lacking \cite{bennett0,flesia,koo,barkema1,trovato,prellberg,barkema2}, 
but it has been argued \cite{bennett} that this is not conclusive since 
much longer SAW's ($10^9$ steps) would be needed to refute the prediction 
unambiguously. As concerns the Manhattan lattice, the numerical evidence 
is unclear since non-universality was found in \cite{bennett}, but 
universality was found in \cite{causo}.

The main question at stake here is the density of parallel contacts (i.e. 
the number of pairs of parallel steps on opposite sites of one plaquette) in a 
very long SAW. It is easily seen that parallel contacts are forbidden for 
2-d self-avoiding closed loops. Therefore, it seems plausible that parallel 
contacts result only from spirals formed by the ends of the walks. This 
was indeed suggested by simulations in \cite{barkema1}, but again it 
seems difficult to perform significant measurements on planar 2-d SAW's.

As pointed out in \cite{bennett}, the natural geometry to study this problem 
is obtained by mapping the plane onto the surface of a cylinder by 
the conformal map 
\be 
   z = x+iy \to w = (L/2\pi) \ln z.        \label{map}
\ee 
Except near the 
ends of the walk, parallel contacts can only occur when the walk wraps 
around the cylinder. In the limit of 
very long chains ($N^\nu \gg L$, where $N$ is the number of steps) 
a typical SAW has to grow either parallel or antiparallel to the cylinder 
axis. Finite size scaling predicts that its longitudinal size $h$ 
(defined here as the average end-to-end distance, projected onto the 
direction parallel to the cylinder axis) scales as 
\be
   h \sim N / L^{1/\nu-1}, \qquad \nu = 3/4 \;.     \label{h}
\ee
The scaling of the winding number variance follows from the facts that 
it must be linear in $h$, and that it is dimensionless, 
\be
   \langle W^2\rangle \approx A \;h/L                 \label{W}
\ee
where $A$ should be a universal amplitude. In the following, winding 
numbers will be measured in terms of wrappings around the cylinder, with 
$W=1$ corresponding one full turn. Finally, the connectivity constant (the 
critical monomer fugacity) should show the usual finite size behavior 
\be
   \mu(L) = \mu_\infty - b\;L^{-1/\nu} .              \label{mu}
\ee
Here, the amplitude $b$ is not universal. A universal amplitude can be 
obtained for the free energy $F = -\ln Z \approx N\;\ln \mu$, where $Z$ 
is the partition sum. Using eq.(\ref{h}) we obtain 
\be
   F(L,N) = F_\infty(N) - B\;h/L,                     \label{F}
\ee
up to terms logarithmic in $N$.

As discussed in \cite{bennett}, the predictions of \cite{cardy} can be 
understood intuitively from the assumption that the 
density of parallel contacts follows essentially 
the winding number: If the walk wraps once around the cylinder, there 
should be $O(1)$ parallel contacts. Thus the prediction for the average 
number of parallel contacts is  
\be
   n_\| \approx C\; h/L \;,                           \label{n_parall}
\ee
with $C$ being yet another universal amplitude.

Finally, we can rewrite eq.(\ref{h}) in terms of the Flory radius $R$ 
of SAW's in planar geometry. More precisely, we define $R$ as the rms. end-to-end
distance of a free SAW of $N$ steps, $R\sim N^\nu$. Then we obtain 
\be
   {h\over L} = D \left({R\over L}\right)^{1/\nu}       \label{h-R}
\ee
with $D$ being also universal. Using this, we can eliminate $h/L$ from 
eqs.(\ref{W}),(\ref{F}), and (\ref{n_parall}) to obtain 
$(\langle W^2\rangle,\; F_\infty(N)-F(L,N),\;n_\|) = (A',B',C')\;(R/L)^{1/\nu}$,
with universal amplitudes $A'=AD, \; B'=BD,$ and $C'=CD$.

The main objective of the present paper was to test eqs.(\ref{h}-\ref{h-R}) by 
means of Monte Carlo simulations. Since we want to simulate lattices with large 
perimeters $L$, and with $N^\nu \gg L$ (we went up to $L=128$ and $N=60,000$), 
we need a fast algorithm. For SAW's in planar geometry the fastest known algorithm 
is the pivot algorithm (at least if one is not interested in $\mu$). But it is easy 
to see that the pivot algorithm doesn't work well on the cylinder (for sufficiently 
small $L$ one can prove that it isn't even ergodic). We therefore used the 
pruned-enriched Rosenbluth method (PERM) \cite{perm} with markovian anticipation 
\cite{review,markov}. In PERM one starts off by growing chains according to the 
well known Rosenbluth method \cite{madras}, but when their weights become too
large resp. too small one interferes by cloning resp. pruning. Weights have 
to be used since Rosenbluth sampling is biased, and the weights are needed to 
compensate the bias. 

In $k$-step markovian anticipation one uses an additional bias 
based on the statistics of sequences of $k+1$ successive steps. Let us number
the $2d$ directions on a $d$-dimensional (hyper-)cubic lattice as $s=0,\ldots 2d-1$.
A sequence of $k+1$ steps is then encoded as
${\bf S} = (s_{-k},\ldots s_0) \equiv ({\bf s},s_0)$.
By $P_{N,m}({\bf S})$ we denote the statistical weight of all $N$-step chains
in an unbiased sample which
had followed the sequence ${\bf S}$ during steps $N-m-k,\ldots, N-m$. This
can be estimated either in a previous test run or during the present run. The ideal
bias in $k$-step markovian anticipation is 
\be
   p(s_0|{\bf s}) = P_{N,m}({\bf s},s_0) / \sum_{s_0'=0}^{2d-1} P_{N,m}({\bf s},s_0')    \label{p}
\ee
with $N\gg m\gg 1$ (we used $m=150$, and averaged $P_{N,m}({\bf S})$ over all $N>300$).
Thus a step $s_0$ is made more often if this step is anticipated to be more
successful in the far ($m$ steps ahead) future, based on previous experience.
Eq.(\ref{p}) should not be used, of course, for the very first steps of the
chain. There the probabilities $P_{N,m}({\bf S})$ are not appropriate. Our
remedy is {\it ad hoc} but efficient:
When determining the bias for the $n$-th step in a chain, we replaced
$P_{N,m}({\bf S})$ by $P_{N,m}({\bf S})+ const/n$, with $const \approx 20$.
The markovian anticipation bias is of course compensated by a weight factor $\propto
1/p(s_0|{\bf s})$, to guarantee correct sampling.

In our actual simulations we used semi-infinite cylinders. Knowing that the
walks anyhow have to grow either parallel or antiparallel to the 
cylinder axis, we started them at height $=1$ and
put an absorbing barrier at height $=0$, so that they had to grow into the
positive $h$ direction. The ratios on the right hand side of eq.(\ref{p})
depend then on the absolute orientation of the steps and on $L$. 
Obviously the longitudinal bias is larger for
smaller $L$. Most of our simulations were done for the square lattice. 
Results for $L=8$ are shown in fig.1. There, sequences of 9
steps are encoded by integers from 0 to $4^9-1$, with the last step giving
the most significant digits. The upward direction is $s=1$, downward is
$s=3$, and the directions perpendicular to the cylinder axis are 0 and 2.
We see that there is both a strong anisotropy, and a strong dependence on
the shape of the last part of the chain.  The former means that downward
steps are likely to be less efficient, while the latter corresponds to the
fact that a strongly curling walk will have problems to be continued. For
large $L$ we used isotropic markovian anticipation with $k=11$.  It gave
roughly one order of magnitude improvement in speed over plain PERM.  For
small $L$, anisotropic anticipation with $k=8$ gave up to one additional
order of magnitude further improvement.

\begin{figure}[b]
  \begin{center}
    \psfig{file=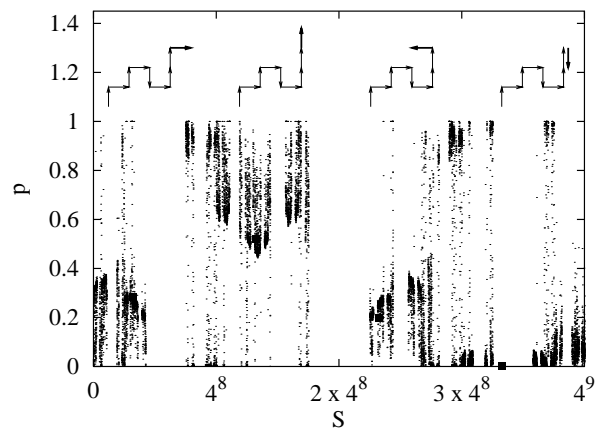,
   width=6.cm,angle=270}
    \vglue0.1cm
    \begin{minipage}{8.5cm}
      \caption{Histogram ratio $p(s_0|{\bf s})$ used for biasing square lattice SAW's 
        on cylinders with $L=8$, based on sequences of length 9. Each sequence ${\bf S}$ 
        is encoded by a number between 0 and $4^9-1 = 262143$. The probability with which
        the next step is taken is $p(s_0|{\bf s})$, up to corrections for the very first
        steps. On all points, statistical errors are $<0.01$. The four
        heavy dots correspond to four sequences ${\bf S}={\bf s}$, ${\bf
          s}+4^8$, ${\bf s}+2\times 4^8$, and ${\bf s}+3\times 4^8$. The chain
        segments corresponding to these sequences are also indicated. The
        probabilities to continue right ({\it r}), up ({\it u}), left ({\it
          l}), and down ({\it d}), after the segment {\it ururdruu}, are
        $p(r):p(u):p(l):p(d) = 0.28:0.52:0.20:0$.}
  \end{minipage}
\end{center}
\label{fig1}
\end{figure}
\vglue-.5cm

We also performed less extensive simulations on the triangular lattice, in 
order to test for universality. There we have 6 local directions instead of 
4, so that we could not use as long memories. We used only isotropic 
anticipation there, with $k=7$.

In all simulations chain lengths were such that $h/L > 80$.  Results
for the connectivity constant are shown in fig.2 where we plotted
$\mu_\infty - \mu(L)$ with $\mu_\infty = 2.638159$ \cite{guttmann} resp. 
$\mu_\infty = 4.150795$ \cite{triangular} for the square resp.
triangular lattice. We see perfect agreement with the theoretical prediction
indicated by the slope of the straight lines. Equation (\ref{h}) was tested
by first making linear fits in plots of $h$ versus $N$ for fixed $L$ (to
avoid biases from the chain ends), and plotting then the slopes against $L$
on a log-log plot. The result is shown in fig.3; again we see perfect
agreement with the theoretical prediction.  Combining the results from
figs. 2 and 3, we find that the amplitude $B = 0.675 \pm 0.002$ is indeed 
universal (the error is just a rough but conservative estimate, as are also 
the following error estimates).

\begin{figure}
  \begin{center}
    \psfig{file=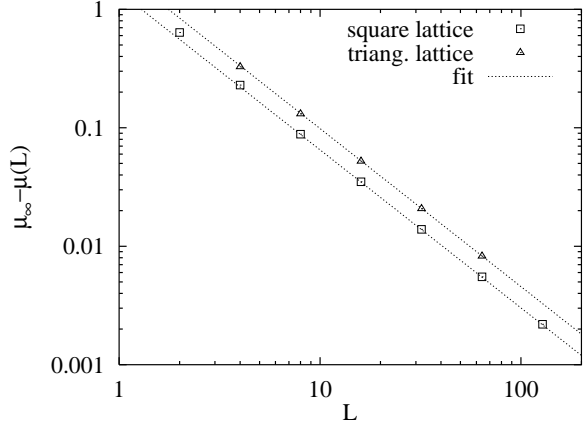,
   width=6.cm,angle=270}
    \vglue0.1cm
    \begin{minipage}{8.5cm}
      \caption{Log-log plot of $\mu_\infty-\mu(L)$ against $L$. The
        straight lines have slope -4/3, as predicted by eq.(\ref{mu}), and
        prefactors adjusted to fit the data.  Error bars on the data points
        are comparable or smaller than the symbol sizes.}
  \end{minipage}
\end{center}
\label{fig2}
\end{figure}
\vglue-5mm
\begin{figure}
  \begin{center}
    \psfig{file=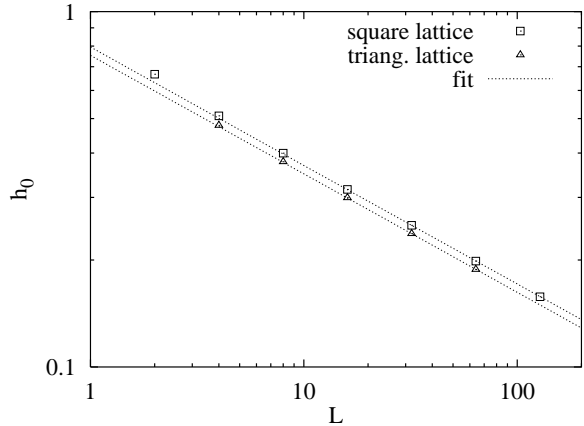,
   width=6.cm,angle=270}
    \vglue0.1cm
    \begin{minipage}{8.5cm}
      \caption{Log-log plot of $h_0$ against $L$, with $h_0$ being the
        constant in a fit $h = h_0N+h_1$ for fixed $L$. Again, error bars
        on the data points are comparable or smaller than the symbol sizes.
        The straight lines have slope $-1/3$ as predicted by
        eq.(\protect{\ref{h}}), with prefactors adjusted to fit the data.}
  \end{minipage}
\end{center}
\label{fig3}
\end{figure}

Winding number variances, multiplied by $L$ and divided by $h$, are shown 
in fig.4. For $L\leq 16$ we see finite size corrections. Apart from these, 
all curves coincide within the expected statistical fluctuations, as is 
expected from eq.(\ref{W}). Again we see that universality is satisfied, 
i.e. the constant $A$ is the 
same for the square and triangular lattices, within the statistical 
errors. More precisely, we find $A = 0.475\pm 0.004$.

In order to estimate the universal amplitude $D$, we use the following
estimates for the end-to-end distances of free chains: $\lim_{N\to\infty}
\langle R^2\rangle /N^{2\nu} = 0.7710\pm 0.0004$ \cite{li} for the square
lattice, and $0.711$ \cite{triangular} for the triangular lattice. Both
gave again consistent results, $D=0.9446\pm 0.0006$.

Finally, we consider the number of parallel contacts (measured only on the 
square lattice). A plot analogous to
fig.4 shows a strong $L$ dependence, but would not allow to estimate this
dependence precisely. More instructive is a plot analogous to fig.3. For
each $L$, we first extracted 
\begin{figure}[hb]
  \begin{center}
    \psfig{file=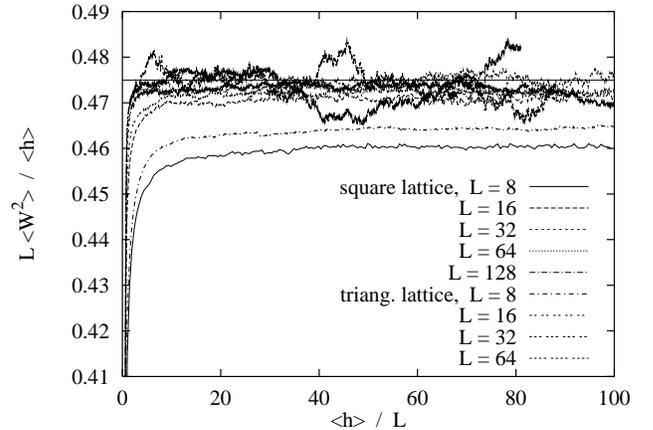,
   width=6.cm,angle=270}
    \vglue0.1cm
    \begin{minipage}{8.5cm}
      \caption{Plot of $L\langle W^2\rangle/h$ against $h/L$. The
        horizontal line is our best estimate for the constant $A$. 
        Statistical errors are typically of the size of the visible 
        fluctuations.} 
  \end{minipage}
\end{center}
\label{fig4}
\end{figure}
\vglue-10mm
\begin{figure}[b]
  \begin{center}
    \psfig{file=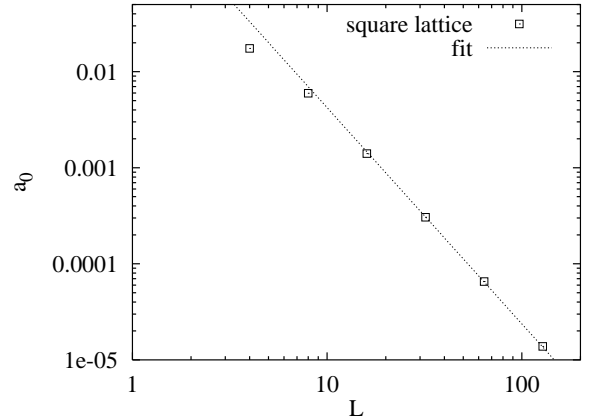,
   width=6.cm,angle=270}
    \vglue0.1cm
    \begin{minipage}{8.5cm}
      \caption{Log-log plot of $a_0$ against $L$, where $n_\| = a_0N+a_1$ 
        for fixed $L$. According to the prediction of
        Ref.\protect\cite{cardy,bennett}, one should have $a_1\sim
        1/L^{1.333}$. The straight line is $const / L^{2.25}$.}
  \end{minipage}
\end{center}
\label{fig5}
\end{figure}

\noindent  the coefficient $a_0$ in a fit $n_\| =
a_0N+a_1$. In fig.5 we then plot $a_0$ against $L$ on double-logarithmic
scale. From this figure we see that $n_\| \sim N L^{-2.25}$ or 
\be 
 n_\|/h \sim L^{-1.92\pm 0.03}.  
\ee 
Again, the error estimate is just an educated
guess.  Indeed, our data show a slight downward curvature, suggesting that
the true exponent might be closer to $-2.0$. In any case, eq.(6) is clearly
ruled out.

Qualitatively this is indeed not surprising. It means that even if the walk
winds once around the cylinder, the chance that it touches itself is much
less than 1 and 
decreases with $L$. This agrees qualitatively with the
behavior of a SAW confined to a strip of width $L$ between 
two repelling walls. In that case the density of monomers at a distance $z$
from a wall increases as $z^{1/\nu}$ \cite{eisen}, implying that the
average distance between two contacts with the wall is $\;\sim L^2.\,$ In
the present case the previously placed part of the SAW acts like a wall,
and during one wrapping around the cylinder the SAW fills a layer of
typical width $L$ in the longitudinal direction. We thus expect in the
present case that the number of parallel contacts per wrapping is $\sim
1/L$.  This argument is of course not rigorous, but it agrees within three
standard deviations with our numerical result.

Finally, let us discuss the universal constant $A$. Naively, one could try to 
estimate it as follows. In planar geometry, it is known that the variance of 
the winding number around either end point is \cite{dupla,prellberg}
\be 
   \langle W^2\rangle = (2\pi^2)^{-1}\ln N\;, \quad N\to\infty  \qquad {\rm (plane)}  \label{WW}
\ee
(notice that we measure windings in units of $2\pi$). When the plane
(punctuated at one of the end points of the chain) is mapped onto the
cylinder by means of eq.(\ref{map}), winding around this end point is
mapped precisely onto wrapping around the cylinder. Taking into account
that $\ln N = const + \nu^{-1} \ln |z| \to const + 2\pi h/\nu L$, we would predict $\langle
W^2\rangle = const + (1/\pi\nu) h/L$, i.e.  $A=1/\pi\nu = 0.4244$. This disagrees
with our measured value by more than 10 standard deviations, and seems
definitely ruled out. To explain this 
discrepancy, we notice that conformal invariance holds only for the canonical
(fixed fugacity $x$) ensemble exactly at the critical point (in this ensemble $N$
fluctuates of course, so that we should actually write
$\langle W^2/\ln N\rangle \to (2\pi^2)^{-1}$ for $x\to x_c\equiv 1/\mu_\infty$,
instead of eq.(\ref{WW})). For any finite $L$, our
results for the cylinder hold, in contrast, for $x_c(L) = 1/\mu(L) > x_c$, since
only there $\langle N\rangle\to \infty$. Therefore,
conformal invariance strictly spoken doesn't give a prediction for $A$.

In summary, we have shown by simulations of very long chains on cylinders 
that universality holds for 2-d SAW's, in contrast to recent claims. 
Our conclusion is based on the fact that parallel contacts are much more 
rare than predicted in \cite{cardy}. This suggests that the corresponding 
operator, which was predicted to be marginal, could play an irrelevant role, 
so that universality should hold for 2-d SAW's. While counting parallel contacts with 
sufficient statistics would have been virtually impossible in planar 
geometry, it is feasible in cylinder geometry. At the same time our results 
on the constant $A$ should serve as a warning that conformal invariance 
should not be applied too naively. 

We thank Gerard Barkema, John Cardy, and Erich Eisenriegler for very useful
discussions, and the latter also for carefully reading the manuscript.

\end{multicols}

\end{document}